\documentclass[epj]{webofc}
\usepackage[utf8]{inputenc}
\usepackage[varg]{txfonts}   % Web of Conferences font
\usepackage{booktabs}
\usepackage{xcolor}
\definecolor{darkred}{rgb}{0.4,0.0,0.0}
\definecolor{darkgreen}{rgb}{0.0,0.4,0.0}
\definecolor{darkblue}{rgb}{0.0,0.0,0.4}
\usepackage[bookmarks,linktocpage,colorlinks,
    linkcolor = darkred,
    urlcolor  = darkblue,
    citecolor = darkgreen]{hyperref}
\wocname{EPJ Web of Conferences}
\woctitle{Lattice2017}
\newcommand{\pref}[1]{(\ref{#1})}
\newcommand{\mf}{\langle x\rangle_{u-d}}
\newcommand{\hm}{\langle x \rangle_{\Delta u - \Delta d}}
\newcommand{\tm}{\langle x \rangle_{\delta u - \delta d}}

\newcommand{\gA}{g_A}

\begin{document}
%%%%%%%%%%%%%%%%%%%%%%%%%%%%%%%%%%%%%%%%%%%%%%%%%%%%%%%%%%%%%%%%%%%%%%%%%%%%%
%
\selectlanguage{english}
%----------------------------------------------------------------------------
\title{%
Multi-hadron-state contamination in nucleon observables from chiral perturbation theory\thanks{Talk given at the 35th International Symposium on Lattice Field Theory, 18 - 24 June 2017, Granada, Spain.}
}
%----------------------------------------------------------------------------
\author{%
\firstname{Oliver} \lastname{B{\"a}r}\inst{1} }
%----------------------------------------------------------------------------
\institute{%
Institut f{\"u}r Physik, Humboldt Universit{\"a}t zu Berlin, Newtonstra{\ss}e 15, D-12489 Berlin, Germany
}
%----------------------------------------------------------------------------
\abstract{%
Multi-particle states with additional pions are expected to be a non-negligi\-ble source of the excited-state contamination in lattice simulations at the physical point. It is shown that baryon chiral perturbation theory  (ChPT) can be employed to calculate the contamination due to two-particle nucleon-pion states in various nucleon observables. Results to leading order are presented for the nucleon axial, tensor and scalar charge and three Mellin moments of parton distribution functions: the average quark momentum fraction, the helicity and the transversity moment. Taking into account experimental and phenomenological results for the charges and moments the impact of the nucleon-pion-states on lattice estimates for these observables can be estimated. The nucleon-pion-state contribution leads to an overestimation of all charges and moments obtained with the plateau method. The overestimation is at the 5-10\% level for source-sink separations of about 2 fm. Existing lattice data is not in conflict with the ChPT predictions, but the comparison suggests that  significantly larger source-sink separations are needed to compute the charges and moments with few-percent precision.  
}
%----------------------------------------------------------------------------
\maketitle
%----------------------------------------------------------------------------
\section{Introduction}	

Chiral perturbation theory (ChPT) is a frequently used tool in the analysis of Lattice Quantum Chromodynamics (QCD) data. Well-known examples are the calculation of finite-volume (FV) effects or the light quark mass dependence of physical observables. ChPT results for the quark mass dependence are commonly used in the chiral extrapolation to relate unphysical lattice results obtained at heavy quark masses to the physical point with quark masses as light as in Nature. However, the need for a chiral extrapolation has been slowly fading away. Constant progress in computer power as well as advances in simulation algorithms have made lattice simulations possible with the light quark masses set to their physical values. Such {\em physical point  simulations} eliminate altogether the systematic uncertainties associated with a chiral extrapolation.

In this talk I briefly report on a different application of ChPT to Lattice QCD, namely the study of excited-state contaminations. In physical point simulations one can expect multi-particle states with additional light pions to become a non-negligible excited-state contamination in many correlation functions measured on the lattice. The calculation of nucleon structure observables is a familiar example where the excited-state contamination is known to be a source of significant systematic uncertainty. The multi-particle states expected to be most relevant in this case are two-particle nucleon-pion ($N\pi$) states, and their contribution is accessible by ChPT.

\section{$N\pi$-state contribution in nucleon correlation function}

Let us start with a simple but illustrative example, the nucleon 2pt function 
\begin{equation} \label{Def2ptfunc}
G_{\rm 2pt}(t) =\sum_{\vec{x}} \, \langle  N(\vec{x},t) \overline{N}(0,0)\rangle\,,
\end{equation}
which is measured in Lattice QCD to compute the nucleon mass. $N,\overline{N}$ denote interpolating fields with the quantum numbers of the nucleon. The sum over the finite spatial volume $V=L^3$ projects to vanishing spatial momentum. Performing the standard spectral decomposition the 2pt function is found to be given by a sum of exponentials,
\begin{equation}
\label{ExpAnsatz}
G_{\rm 2pt}(t) = c_0 e^{-M_N t} + c_1 e^{-E_1 t} + c_2 e^{-E_2 t} + \ldots\,,
\end{equation}
where $t > 0$ and the ordering $M_N < E_1 < E_2 < \ldots $ for the energies is assumed. By construction, the first exponential contains the nucleon mass. The coefficient $c_0$ associated with it is the squared matrix element of the interpolating field between the vacuum and the nucleon at rest. The other terms stem from excited states with the same quantum numbers as the nucleon, and the coefficients $c_j$  involve the matrix elements with the excited states instead of the nucleon state.

In lattice simulations one usually computes the effective nucleon mass, defined as the negative time derivative of $\ln G_{\rm 2pt}(t)$. With \pref{ExpAnsatz} we obtain ($\Delta E_k = E_k-M_N$)
\begin{equation}\label{Meff}
M_{N,{\rm eff}}  = M_N + \frac{c_1}{c_0} \Delta E_1 e^{-\Delta E_{1} t} + \frac{c_2}{c_0} \Delta E_2 e^{-\Delta E_{2} t}+\ldots \,.
\end{equation}
In the limit $t\rightarrow \infty$ the effective mass converges to the nucleon mass. For large but finite $t$ there is an excited-state contamination present.  Although exponentially suppressed the euclidean time separation needs to be sufficiently large for the excited-state contamination to be small. In practice the signal-to-noise problem \cite{Parisi:1983ae,Lepage:1989hd} prevents us from going to very large euclidean time separations to make the excited-state contribution arbitrarily small. The smaller the pion mass the shorter the time separation for which the effective mass can be measured with small statistical errors. Typical time separations at present are between 1 to 1.5 fm.

\begin{figure}[tb]
\centerline{\includegraphics[scale=0.4]{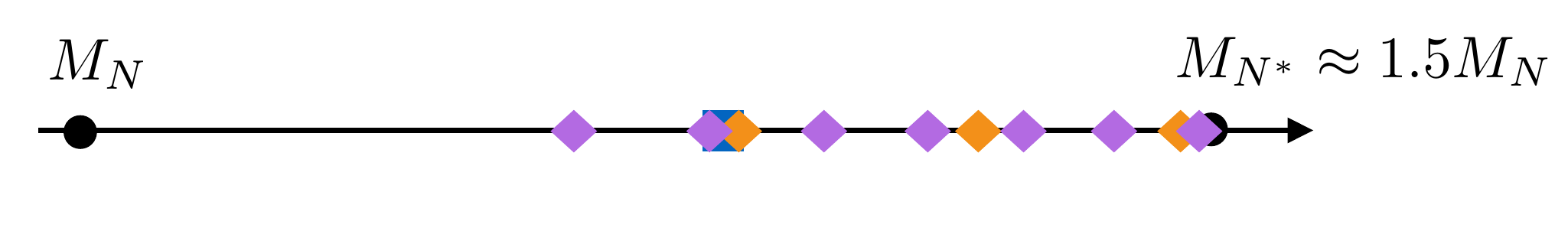}}
\caption{Sketch of the energies of the lowest $N\pi$ states (diamonds) that contribute to the nucleon 2pt function. Discrete momenta corresponding to  $M_{\pi}L=4$ (orange) and $M_{\pi}L=6$ (purple) are considered and interaction energies are ignored. The blue square represents the energy of the $N\pi\pi$ state with all three particles at rest. }
\label{fig:sketch}
\end{figure}

From experiment we can get some idea about the states that contribute to the 2pt function in a finite spatial volume. There is a contribution from resonance states that are associated to the nucleon resonances in infinite volume, the most prominent one being the Roper resonance $N^*(1440)$.\footnote{The precise relation between infinite volume resonance properties and the finite volume energy spectrum has been worked out first in Ref.\ \cite{Luscher:1991cf}.} In addition there is the contribution from multi-particle states, e.g.\ 2-particle $N\pi$ states, 3-particle $N\pi\pi$ states etc. The multi-particle-state contribution is expected to become rapidly relevant  for pion masses approaching the physical value. This is easily seen by ignoring the interaction energy of the particles, which is expected to be rather small since the pions interact only weakly with the nucleon and with themselves. Within this approximation the energy of the $N\pi\pi$ state with all three particles at rest is equal to $M_N+2M_{\pi}\approx 1.3 M_N$. For a two-particle $N\pi$ state to contribute both nucleon and pion need non-vanishing and opposite spatial momenta. If we assume periodic boundary conditions for the finite spatial volume the spatial momenta are discrete, $\vec{p}_k=2\pi\vec{k}/L$, with $\vec{k}$ having integer-valued components. The larger the lattice extent $L$ the smaller the discrete energies of the $N\pi$ states allowed by the boundary conditions. If we assume the typical value $M_{\pi}L=4$ we find three $N\pi$ states with an energy less than $M_{N^*}\approx 1.5M_N$ (see fig.\ \ref{fig:sketch}). This number increases to seven for $M_{\pi}L=6$, as realized in the simulations of the PACS collaboration \cite{Ishikawa:2015rho}.  At this conference Y.~Kuramashi reported on simulations on even larger lattices with $M_{\pi}L\approx 7.7$  \cite{Kuramashi:Lattice17}, and such volumes imply ten $N\pi$ states with energy below $1.5M_N$. The main conclusion is that quite a few multi-particle states are expected to contribute to the sum in eq.\ (\ref{ExpAnsatz}) before a contribution of the first resonance state appears. Thus, in physical point simulations the multi-particle states dominate the excited-state contamination in the asymptotic regime of large but still finite $t$. 

The impact of the multi-particle states to the 2pt function depends also on the size of the coefficients $c_j, j\ge1$. Not much is known  a priori about these coefficients, except for the fact that a coefficient associated with an $n$-particle state is, in a finite spatial volume, suppressed by $(1/L^3)^{n-1}$. Note that this volume suppression cannot be taken as an argument for a small multi-particle-state contribution if $L$ is large. Although the contribution of each individual multi-particle state decreases, more and more states contribute the larger the volume is. In particular, the multi-particle-state contribution is not a FV effect and will not disappear in the infinite-volume limit.  

Other observables than the nucleon mass are expected to be even more afflicted by an excited state contamination. Still rather straightforward to compute numerically are nucleon charges and Mellin moments of parton distribution functions. All these observables refer to nucleon matrix elements $\langle N(\vec{p})|O_X| N(\vec{p})\rangle$ of a local quark bilinear $O_X$ between nucleon states and with zero momentum transfer. In case of the standard axial vector current as well as the tensor and scalar density for $O_X$ ($X=A,T,S$) one obtains the axial, tensor and scalar charges $g_A, g_T$ and $g_S$.\footnote{Throughout I will exclusively refer to the flavor nonsinglet charges.} The axial charge is known at the per mille level form neutron beta decay \cite{Olive:2016xmw} and considered to be a benchmark observable for lattice QCD. The scalar and tensor charge are only poorly known, but enjoy a revived interest. New nuclear beta-decay experiments aim at an order of magnitude more precise upper bounds for these charges. In order to constrain beyond-standard-model physics lattice QCD estimates for these charges with 10-15\% uncertainties will be needed \cite{Bhattacharya:2011qm}.

Parton distribution functions (pdfs) are difficult to access directly in Euclidean space-time. Instead, one can define and calculate Mellin moments that can be related to matrix elements of local one-derivative operators. For example, the average quark momentum fraction $\mf$ (also called the first unpolarized moment) involves the operator $ V_{\mu\nu} \,=\, \overline{u}\gamma_{{\{\mu}}D^-_{\nu\}} u - \overline{d}\gamma_{{\{\mu}}D^-_{\nu\}} d$ with the (color covariant) derivative  given by $D^-_{\mu} = ( {\overrightarrow D}_{\mu} -  \overleftarrow{D}_{\mu})/2$, and the curly brackets refer to symmetrization with respect to the indices $\mu,\nu$ and subtraction of the trace. The helicity moment $\hm$ and the transversity moment $\tm$ are defined analogously and differ in the gamma matrices involved. Pdf moments are accessible phenomenologically and some of them have been determined with percent accuracy. For example, in Ref.\ \cite{Alekhin:2012ig} the average quark momentum fraction is found as $\langle x\rangle_{u-d,{\rm phen}}=0.1655(39)$ (in the $\overline{\rm MS}$ scheme at 2 GeV).

The standard methods to compute the charges and moments require the computation of the 3pt function.\footnote{The recently proposed method \cite{Bouchard:2016heu} inspired by the Feynman-Hellman theorem avoids the explicit calculation of the 3pt function.} To be specific, we need the 3pt function 
\begin{equation}
G_{{\rm 3pt},A}(t,t') = \sum_{\vec{x},\vec{y}} \,\Gamma'_{k,\alpha\beta} \langle N_{\beta}(\vec{x},t) A^3_k(\vec{y},t') \overline{N}_{\alpha}(\vec{0},0)\rangle\,
\end{equation}
in case of the axial charge. The nucleon interpolating fields are the same as in the 2pt function. A spatial component of the non-singlet axial vector current is placed at some operator insertion time $t'$ between source and sink, and the projector $\Gamma'_{k}$ is chosen such that the leading part of the 3pt function is proportional to $g_A$. Forming the ratio with the 2pt function,
\begin{equation}\label{Defratiogeneric}
R_A(t,t')=\frac{G_{{\rm 3pt},A}(t,t')}{G_{\rm 2pt}(t)}\,,
\end{equation}
and performing the spectral decomposition of both numerator and denominator one finds the representation 
\begin{equation}
\label{DefRatio}
R_A(t,t')= g_A + b_{A,1} e^{-\Delta E_1 (t-t')} + \tilde{b}_{A,1} e^{-\Delta E_1 t'} + \tilde{c}_{A,1} e^{-\Delta E_1 t } +\ldots\,
\end{equation}
for the ratio. In the limit that all times $t,t'$ and $t-t'$ go to infinity the ratio tends to a constant, the axial charge. For finite times there are again excited-state contributions present stemming from resonance and multi-hadron states. The coefficients $b_{A,1},\tilde{b}_{A,1},\tilde{c}_{A,1}$ in \pref{DefRatio} are ratios of  matrix elements involving the axial vector current as well as the interpolating fields and the states for the vacuum, the nucleon at rest and the first excited state.

The excited-state contributions in eq.\ \pref{DefRatio} involves the three time separations $t-t', t'$ and $t$. As already mentioned, the source-sink separations $t$ accessible in numerical simulations are limited due to the signal-to-noise problem. Since $t-t'$ and $t'$ are even smaller than $t$ we anticipate a significantly larger excited-state contamination in calculations of the axial charge (as well as the other charges and the pdf moments). For large time separations, however, we expect the multi-hadron states to be the dominant source for the excited-state contribution, since these have the smallest gap to the energy of the single nucleon state. 

Topic of this talk is that the correlation function ratios can be computed in ChPT. Such calculations provide the coefficients  $b_{A,j},   \tilde{b}_{A,j},\tilde{c}_{A,j} $ in \pref{DefRatio} associated with multi-particle-state contributions. Results are available for the two-particle $N\pi$ contribution in the ratios for the three nucleon charges and the three pdf moments introduced before \cite{Bar:2015zwa,Bar:2016uoj,Bar:2016jof}. For a recent review see Ref.\ \cite{Bar:2017kxh}. Calculations for contributions  of the  $N\pi\pi$ and $\Delta\pi$ states will be analogous but have not been performed yet.

That BChPT can be employed to compute the multi-particle state contribution in correlation functions is neither surprising nor a new idea \cite{Tiburzi:2009zp,Bar:2012ce}. In fact, 
Brian Tiburzi \cite{Tiburzi:2009zp} was probably the first to compute the $N\pi$-state contribution to $R_A$ in an attempt to explain, at least qualitatively, why most lattice results at the time underestimated the experimental value of the nucleon axial charge. However,  the $N\pi$-state contribution is found to lead to an overestimation of the axial charge and the idea did not receive much attention.

In the following I will present results for the nucleon axial, tensor and scalar charges and the three pdf moments introduced before. The calculation differs in some details from the early one in Ref.\ \cite{Tiburzi:2009zp}. The discreteness of the nucleon and pion momenta due to a finite spatial volume is taken into account. In addition, the mapping of smeared nucleon interpolating fields, commonly used in lattice calculations, to ChPT has been put on firmer grounds \cite{Luscher:2013vga,Bar:2013ora}. More importantly, having results for six instead of only one observable we find the $N\pi$-state contribution in all six observables to be related. This is nothing but chiral symmetry at work.

\section{The correlation functions in ChPT}

Chiral perturbation theory refers to the low-energy effective theory of QCD based on spontaneous chiral symmetry breaking \cite{Weinberg:1978kz,Gasser:1983yg,Gasser:1984gg}. It has been around for many years and there seems no need to introduce or review it here.\footnote{There exist numerous reviews and lecture notes on ChPT that differ in scope and length, e.g.\ Refs.\ \cite{Colangelo:2000zw,Gasser:2003cg,Kubis:2007iy,Scherer:2009bt,Golterman:2009kw,Scherer:2012xha,Ecker:2013xja}. A particularly useful review of Baryon ChPT that also covers applications to lattice QCD is given in Ref.\ \cite{Bernard:2007zu}.} 

The two most prominent applications in Lattice QCD are the calculation of the quark mass dependence of observables and FV effects due to the pions. Starting point for the application we have in mind here, the excited-state contribution caused by two-particle $N\pi$ states, is the same well-established formulation of Baryon ChPT (BChPT) \cite{Gasser:1987rb,Becher:1999he} that is basis for the other applications as well. We work to leading order only, thus we do not encounter any of the subtleties associated with the loop expansion in BChPT. We consider SU(2) BChPT and assume isospin symmetry. In this case the chiral effective theory contains the three mass degenerate pions $\pi^a$ and the nucleon doublet $\Psi=(p,n)^T$ involving the mass degenerate proton and neutron fields. The interaction term relevant here is given by\footnote{We work in Euclidean space time.} 
\begin{equation}
{\cal L}_{{\rm int},1\pi} =\frac{ig_A}{2f}\overline{\Psi}\gamma_{\mu}\gamma_5\sigma^a\Psi \partial_{\mu}\pi^a\,,
\end{equation}
which implies the well-known one-pion-exchange-potential (OPEP) between a nucleon pair.  It is proportional to the ratio $g_A/f$ of two LO low-energy coefficients (LECs), the chiral limit values of the axial charge and the pion decay constant. 

The effective operators for the axial vector current and the scalar density  have been known for a long time \cite{Gasser:1987rb,Fettes:2000gb}. The operators needed for the average quark momentum fraction and the helicity moment too can be taken from the literature \cite{Dorati:2007bk,Wein:2014wma}. The expressions associated to the tensor operators, needed for the tensor charge and the transversity moment, have been derived following the general procedure described in Ref.\ \cite{Fettes:2000gb}. The explicit expressions are not displayed here but can be found in Refs.\ \cite{Bar:2016uoj,Bar:2016jof}. Still, it is worth pointing out that the chiral limit values of all charges and pdf moments we are interested in appear as LECs in these effective operators. These are input parameters of ChPT. 

The chiral expressions for the nucleon interpolating fields are also needed. These too are derived based on the transformation properties under chiral symmetry and parity \cite{Nagata:2008zzc}. The standard local 3-quark baryon operators \cite{Ioffe:1981kw,Espriu:1983hu} have been mapped to ChPT in Ref.\ \cite{Wein:2011ix}. Smeared interpolating fields are mapped to the same pointlike expressions but with different LECs provided the smeared quark fields transform as their local analogues under parity and chiral symmetry. This, for instance, is the case in Gaussian and exponential smearing \cite{Gusken:1989ad,Gusken:1989qx,Alexandrou:1990dq} or the gradient flow \cite{Luscher:2013cpa}. However, the smearing radius must be small compared to the Compton wavelength of the pion, i.e.\ $R_{\rm smear} \ll 1/M_{\pi}$ \cite{Luscher:2013vga,Bar:2013ora}. For physical pion masses this inequality seems reasonably well satisfied for smearing radii up to a few tenths of a fermi. 

\begin{figure}[tb]
\centering
\includegraphics[scale=0.45]{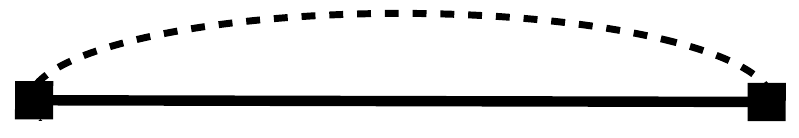}\hspace{1cm}\includegraphics[scale=0.45]{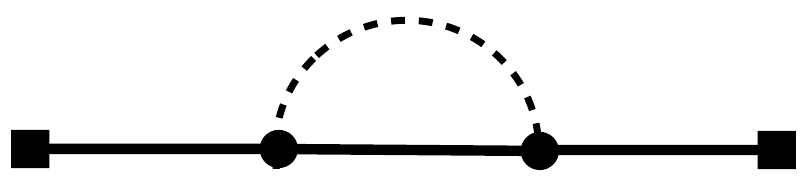}\\
a)\hspace{4.4cm} b)\\[3ex]
\includegraphics[scale=0.45]{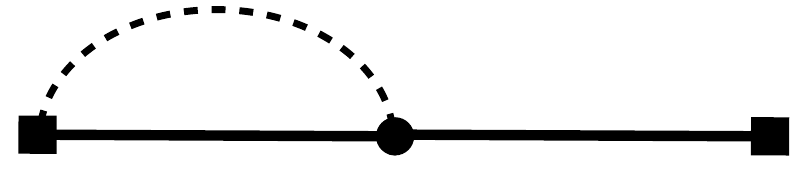}\hspace{1cm}\includegraphics[scale=0.45]{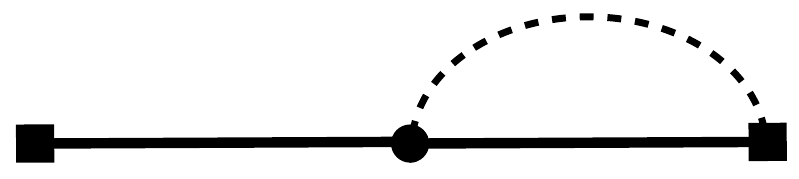}\\
c)\hspace{4.4cm} d)
\caption{Feynman diagrams for the nucleon 2pt function. The squares represent the nucleon interpolating fields at times $t$ and $0$. The circles represent a vertex insertion at an intermediate space-time point; and an integration over this point is implicitly assumed. The solid and dashed lines represent nucleon and pion propagators, respectively. }
\label{fig:Npidiagrams2pt}
\end{figure}

\begin{figure}[t]
\centering
\includegraphics[scale=0.45]{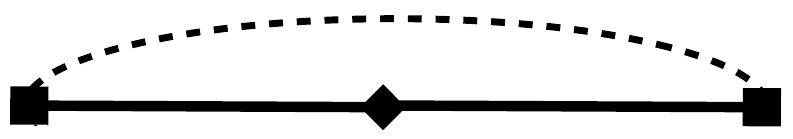}\hspace{0.3cm}\includegraphics[scale=0.45]{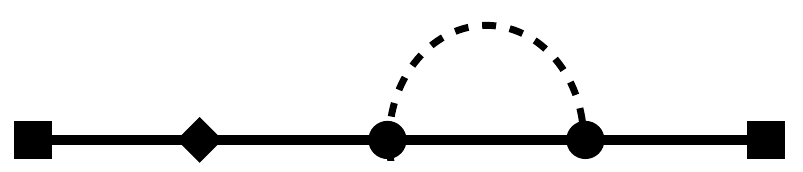}\hspace{0.3cm}\includegraphics[scale=0.45]{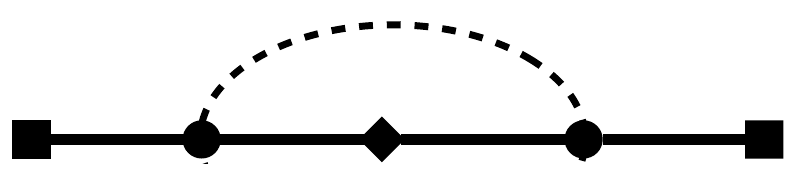}\\[4ex] 
\includegraphics[scale=0.45]{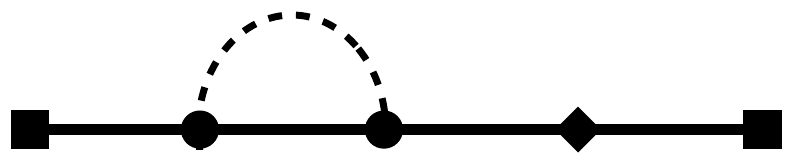}\hspace{0.3cm} \includegraphics[scale=0.45]{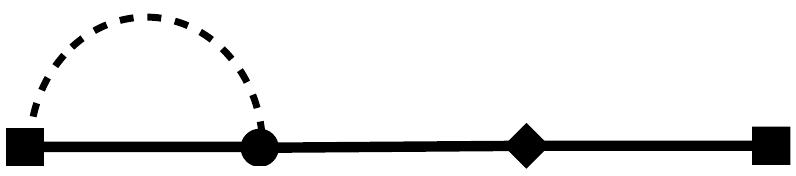}\hspace{0.3cm}\includegraphics[scale=0.45]{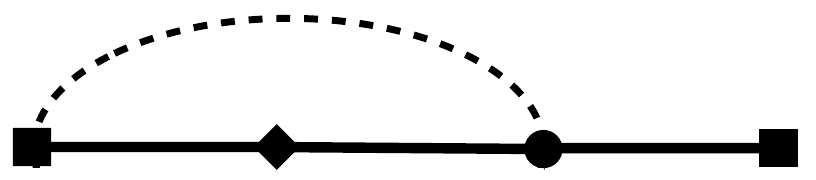}\\[4ex] 
\includegraphics[scale=0.45]{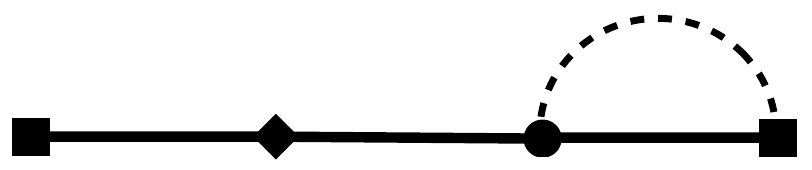}\hspace{0.3cm}\includegraphics[scale=0.45]{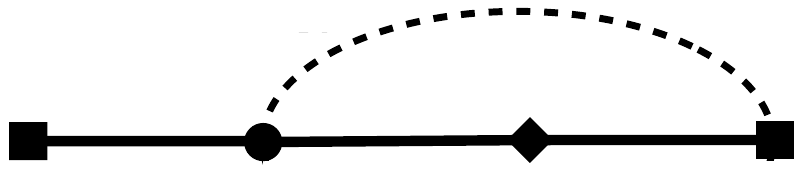}\hspace{0.3cm}
\includegraphics[scale=0.45]{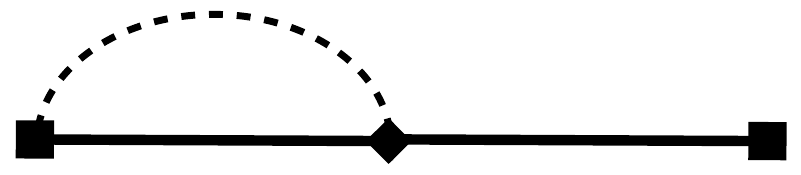}\\[4ex]  
\includegraphics[scale=0.45]{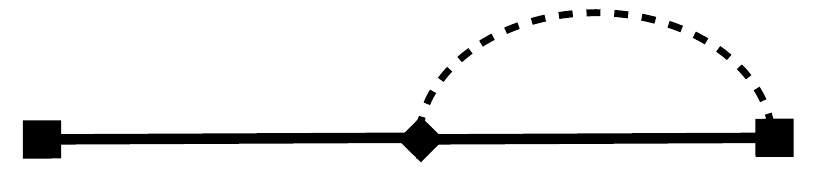}\hspace{0.3cm}\includegraphics[scale=0.45]{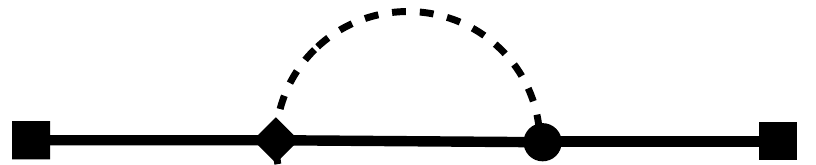}\hspace{0.3cm}\includegraphics[scale=0.45]{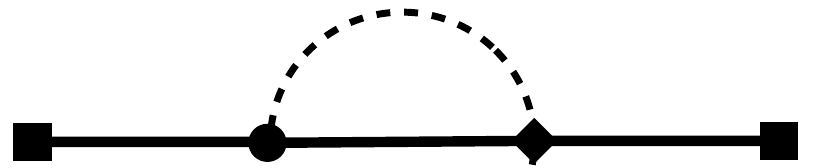}\\[4ex] 
\caption{Feynman diagrams for the LO nucleon-pion contribution in the 3pt functions. The diamond stands for the operator insertion at time $t'$.}
\label{fig:Npidiagrams3pt}
\end{figure}

With these elements the calculation of the 2pt and 3pt functions is a standard perturbative calculation in ChPT. It seems convenient to use the time-momentum representation for the pion and nucleon propagators since it directly yields the time dependence of the correlation functions. 
Figure \ref{fig:Npidiagrams2pt} shows the leading diagrams with an $N\pi$-state contribution to the 2pt function. These diagrams contain a contribution that drops off exponentially with the total energy $E_{N\pi,n} = E_{N,n}+E_{\pi,n}$ of a nucleon-pion state, where the discrete spatial momenta are back-to-back, $\vec{p}_{N,n}=-\vec{p}_{\pi,n}$. We are interested in the prefactor of this exponential. Note that even though the diagrams in fig.\ \ref{fig:Npidiagrams2pt} are 1-loop diagrams the calculation of the $N\pi$-state contribution does not involve a summation over some undetermined momentum, i.e.\ it is a tree-level calculation.\footnote{Diagrams b) - d) also contain single-nucleon contributions dropping off with $\exp(-M_N t)$, and this contribution does involve a sum over all momenta of the intermediate $N\pi$ pair. Diagram b), for instance, is essentially the nucleon self-energy diagram that leads to the well-known $M_{\pi}^3$ term in the 1-loop ChPT result for the nucleon mass \cite{Gasser:1987rb}.}
Figure \ref{fig:Npidiagrams3pt} shows the leading diagrams with an $N\pi$-state contribution to the 3pt functions. 

The calculation of the Feynman diagrams is straightforward. The result for the ratio of the 3pt and 2pt functions is of the general form
\begin{equation}
\label{DefRatio2}
R_X(t,t')= g_X \Big[1+ \sum_{\vec{p}_n} \Big(b_{X,n} e^{-\Delta E_n (t-t')} + \tilde{b}_{X,n} e^{-\Delta E_n t'} + \tilde{c}_{X,n} e^{-\Delta E_n t }\Big)\Big]
\end{equation}
for all three nucleon charges ($X=A,T,S$), with $\Delta E_n = E_{N\pi,n}-M_N$. An analogous result is found for the moments. The non-trivial results of the calculation are the coefficients $b_{X,n}, \tilde{b}_{X,n}, \tilde{c}_{X,n}$ in all six ratios. The exact expressions for these coefficients can be found in Refs.\ \cite{Bar:2016uoj,Bar:2016jof} and are not displayed here. Still, it is noteworthy that all coefficients involve a product of two universal factors, for example 
\begin{equation}
 \tilde{c}_{{X},n} = \frac{1}{16(fL)^2E_{\pi,n}L} \left(1-\frac{M_N}{E_{N,n}}\right)\Bigg(\ldots\Bigg)\,.
\end{equation}
The first factor on the right hand side shows the anticipated volume suppression factor $1/L^3$ of a two-particle state in a finite spatial volume. It combines with the pion decay constant and the pion energy to the dimensionless combination $1/(fL)^2E_{\pi,n}L$. The second factor $(1-M_N/E_{N,n})$ vanishes if the momentum of the nucleon (and the pion) is zero. This has to be the case since the nucleon-pion state with both particles at rest does not contribute to the correlation function because it is parity odd. The remaining factor is non-universal and not very illuminating. 

More interesting than the exact expressions is the fact that the coefficients do not depend on the LECs associated with the nucleon interpolating fields. To LO these cancel in the ratio $R_X$. Consequently, the LO $N\pi$-state contribution to the ratios is the same for pointlike and for smeared interpolating fields, since the difference between these interpolators is encoded in their different values for the LECs. 
However, this universality property will be lost at higher orders in the chiral expansion.

The coefficients depend on a few parameters only: The pion mass and pion decay constant, the finite volume via $M_{\pi}L$ and four more LECs (the chiral limit values of the nucleon mass, the axial charge, the average quark momentum fraction and the helicity moment). All these parameters can be estimated using their experimental or phenomenological values. With this input we can estimate the $N\pi$-state contribution to the ratios calculated in lattice simulations. 

\section{Impact on lattice calculations}
The simplest estimator for the charges and pdf moments is the so-called midpoint estimate. It is based on a simple observation: For a given source-sink separation $t$ the excited-state contribution to $R_X$ is minimized if the operator insertion time $t'$ is in the middle between source and sink. Therefore, the best estimate for the charges and moments is the midpoint value $R_X(t,t/2)$. It is essentially equivalent to what is called plateau estimate, and we will use this terminology in the following. 

The experimental/phenomenological values for the masses and LECs can be found in the literature \cite{Olive:2016xmw,Alekhin:2012ig,Blumlein:2010rn}.\footnote{For the results presented here the following simplified values have been used: $M_{\pi}=140$ MeV, $M_N=940$ MeV, $\gA=1.27$, $f_{\pi}= 93$ MeV, $\mf=0.165$ and $\hm=0.19$. Errors in these estimates are ignored since they are too small to be significant for the LO ChPT results.} With these input parameters being fixed the LO ChPT result for  $N\pi$-state contribution is a function of the spatial volume only.

\begin{figure}[t]
\begin{center}
$R_A(t,t/2)/g_A$\\
\includegraphics[scale=0.85]{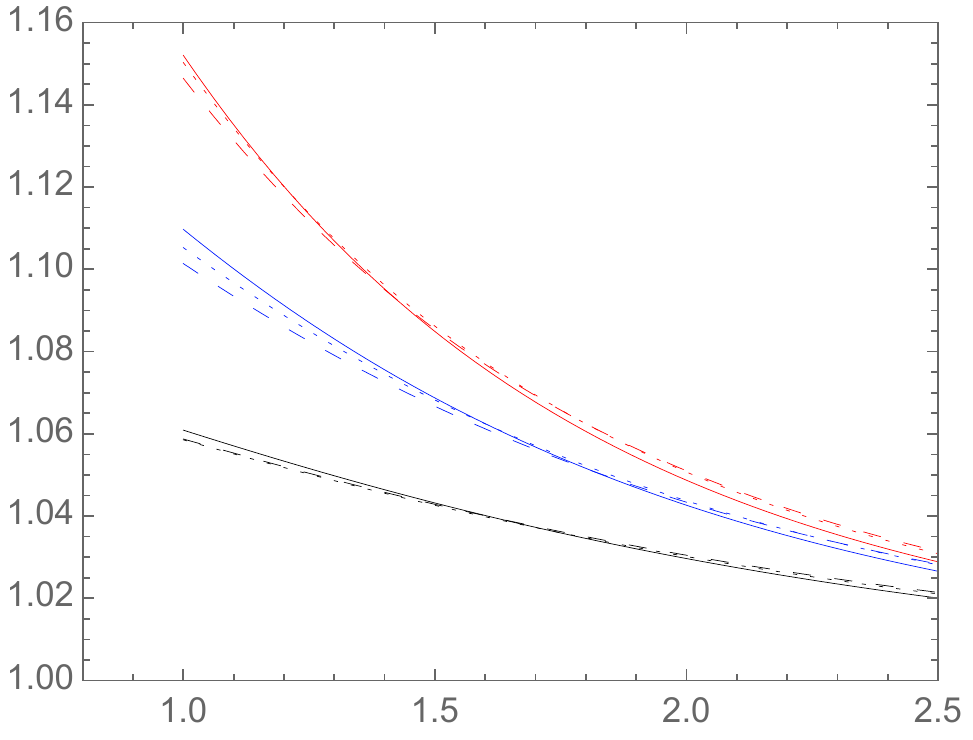}\\
$t$ [fm]
\caption{The plateau estimate $R_A(t,t/2)$ normalized by $g_A$ as a function of the source-sink separation $t$. Results are shown for $M_{\pi}L=4$ (solid lines),  $M_{\pi}L=5$ (dotted lines) and $M_{\pi}L=6$ (dashed lines). Different colors distinguish between three different upper energy bounds for the $N\pi$ states taken into account (see main text).}
\label{fig:axialcharge}
\end{center}
\end{figure}

Fig.\ \ref{fig:axialcharge} shows the plateau estimate $g_{A,{\rm plat}}(t)\equiv R_A(t,t/2)$ for the axial charge divided by $g_A$ 
 as a function of the source-sink separation $t$. 
Without the excited-state contribution this ratio would be equal to 1. Any deviation from this constant value shows directly the $N\pi$ contribution to the plateau estimate in percent. Results are shown for three different lattice sizes with $M_{\pi}=4,5$ and 6 (solid, dotted and dashed lines).  Different colors refer to different numbers of $N\pi$ states taken into account in the sum over the discrete spatial momenta $\vec{p}_n$, cf.\ equation \pref{DefRatio2}. For the black lines $N\pi$ states with a total energy up about 1.35 GeV are included. The blue and red lines correspond to total energies up to about 1.6 and 1.9 GeV.
Note that the number of states corresponding to these energies depends on the spatial lattice extent $L$. For example, the black lines in fig.\ \ref{fig:axialcharge} 
correspond to 2, 3 and 5 states for $M_{\pi}L=4,5$ and 6, respectively. 

Apparently, the differences between lines of the same color are very small, i.e.\ the $N\pi$ contribution is essentially the same for the three different $M_{\pi}L$ values displayed in the figure. As already mentioned in the introduction, the $1/L^3$ suppression of the two-particle-state matrix elements is expected to be largely compensated by a growing number of $N\pi$ states in a given energy interval. Still, it is perhaps somewhat surprising that the differences between the three different volumes are so small.

Fig.\ \ref{fig:axialcharge} also shows that the $N\pi$ contribution leads to an overestimation of the axial charge by the plateau estimate. Before reading off a number from the plot recall that ChPT is an expansion in small pion momenta. We may expect reasonably reliable ChPT results for the low-momentum $N\pi$ states, depicted by the black lines in the plot. The chiral expansion of the high-momentum $N\pi$ states, included in the blue and red curves, is expected to work less well if it works at all. 

This restriction implies that the source-sink separation $t$ needs to be large enough for the contribution of the high-momentum states to be sufficiently suppressed such that the low-momentum states provide the dominate part of the total $N\pi$ contribution. According to fig.\ \ref{fig:axialcharge} this seems satisfied for $t\approx 2.5$ fm, but certainly not for $t\approx 1.5$ fm where the low-momentum states contribute less than half of the contribution given by the red curves. It is at about $t\approx 2$ fm that the black lines start to capture the dominant $N\pi$ contribution, thus we may conclude that source-sink separations of about 2 fm and larger are required for ChPT to make reliable estimates. How large the higher order corrections at these source-sink separations are is difficult to predict. Naive error estimates suggest a 30-50\% uncertainty for $t\gtrsim2$ fm. A more solid error estimate requires the calculation at NLO. 

\begin{figure}[t]
\centering$R_X(t,t/2)/g_X$ and $R_X(t,t/2)/\Pi_X$ \\
\includegraphics[scale=0.6]{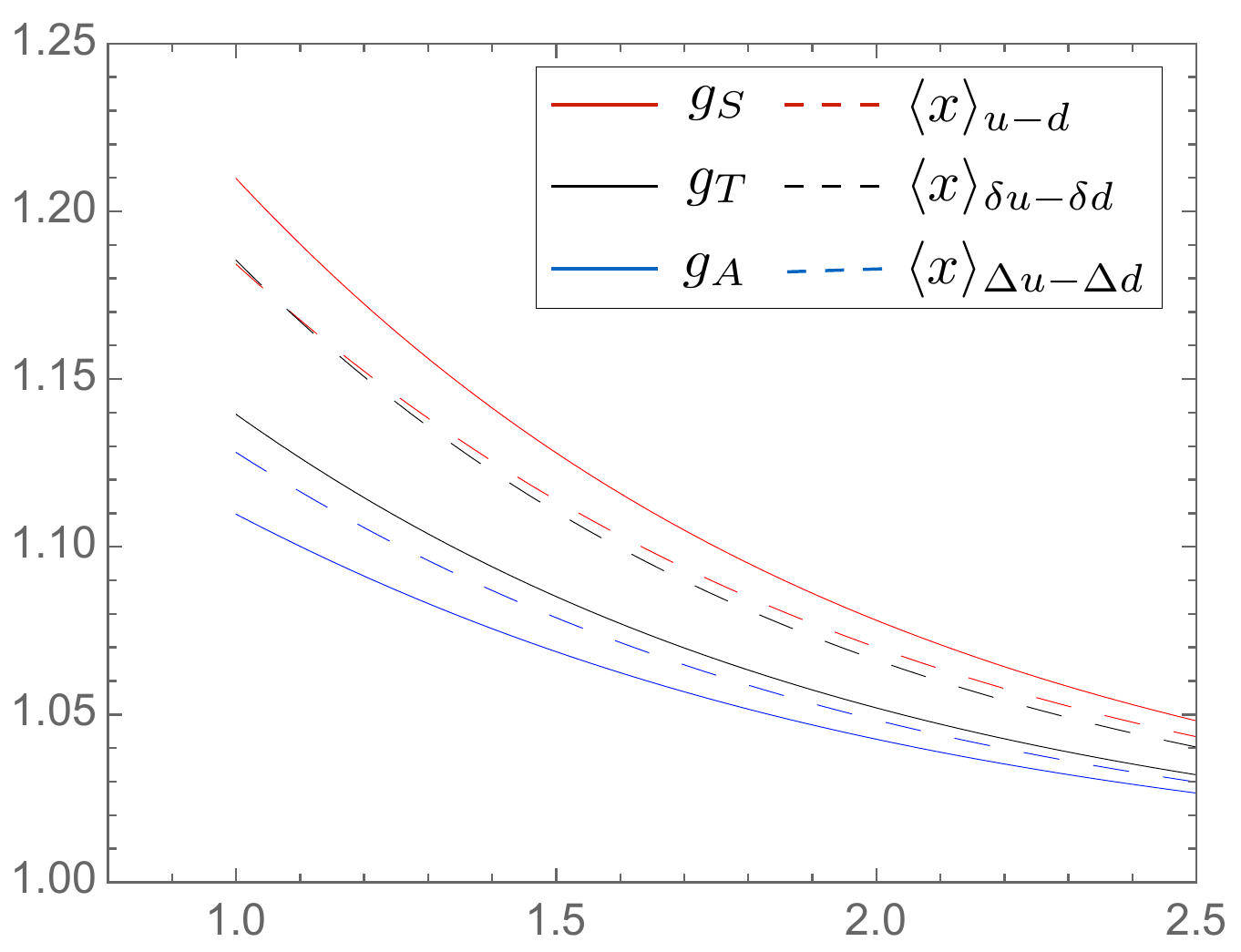}\\
$t$ [fm]
\caption{The plateau estimates $R_X(t,t/2)$ normalized by the asymptotic values for all charges and moments. Results are shown for $M_{\pi}L=4$ with $N\pi$ states included with energies up to 1.6 GeV. The result for the axial charge (blue solid line) is the same as the blue solid line in fig.\ \ref{fig:axialcharge}.}
\label{fig:resultsall}
\end{figure}

Qualitatively the same results are found for the tensor and scalar charges and the three moments. Fig.\ \ref{fig:resultsall} shows the plateau estimates normalized by either the charge or the moment. As for the axial charge we only find a small dependence on $M_{\pi}L$ in all cases, 
so fig.\ \ref{fig:resultsall}  shows the $M_{\pi}L=4$ results only. In all cases we find the low-momentum $N\pi$ states start to dominate for $t\gtrsim 2$ fm, thus the results displayed in fig.\ \ref{fig:resultsall} are expected to be reliable only for $t\gtrsim2$ fm.

According to fig.\ \ref{fig:resultsall} the $N\pi$-state contribution leads to an overestimation for all three charges and the three moments by their plateau estimates. The largest overestimation is found for the scalar charge, which is about twice as large as for the axial charge. The $N\pi$ contribution to the average momentum fraction is close to the one for the scalar charge. In summary we find an overestimation of about 5-10\% for the charges and moments at $t\approx 2$ fm, and it slowly decreases to 3-6\% at $t\approx 2.5$ fm. 

We stress once again that these numbers are results at LO, i.e.\ at O($p$).  The higher order corrections are hard to quantify without having done the calculation at NLO. However, naive arguments can be made for the O($p^2$) corrections and suggest error estimates up to about 50\%.  

\section{Comparison with lattice data}

Most of the existing lattice results for the charges and moments have been obtained for pion masses larger than the physical value. There exist many reviews summarizing these results, for example Refs.\  \cite{Syritsyn:2014saa,Green:2014vxa,Constantinou:2015agp,Alexandrou:2016hiy}. Moreover, Sara Collins presented a review talk at Lattice 2016 in Southampton \cite{Collins:Lattice16}, and I refer to it for details concerning the present status of the lattice calculations.  

Some collaborations have carried out lattice calculations with the pion mass at or near the physical value \cite{Bali:2014nma,Abdel-Rehim:2015owa,Bhattacharya:2016zcn,vonHippel:2016wid,Green:2012ud,Bali:2014gha}. The main obstacle for directly applying the ChPT results to these lattice calculations are the rather small source-sink separations in these simulations. In most cases the maximal source-sink separation  used to extract the charges and moments with the plateau method is below 1.5 fm. As discussed in the previous section, ChPT is not expected to provide solid results unless the source-sink separations are about 2 fm or even larger. Still, it is useful to compare the existing lattice results with our ChPT predictions, since the latter need to be reproduced eventually.

In the following we restrict ourselves to data obtained with the plateau method with pion masses not larger than 165 MeV. Lattice results obtained with the summation method \cite{Green:2012ud}, two-state fits \cite{Bhattacharya:2016zcn} or the method based on the Feynman-Hellman theorem \cite{Bouchard:2016heu,Berkowitz:2017gql} cannot be used in our comparison.

\begin{figure}[t]
\centering
$g_{A,{\rm plat}}(t)/g_{A,{\rm exp}}$ \\
\includegraphics[scale=0.85]{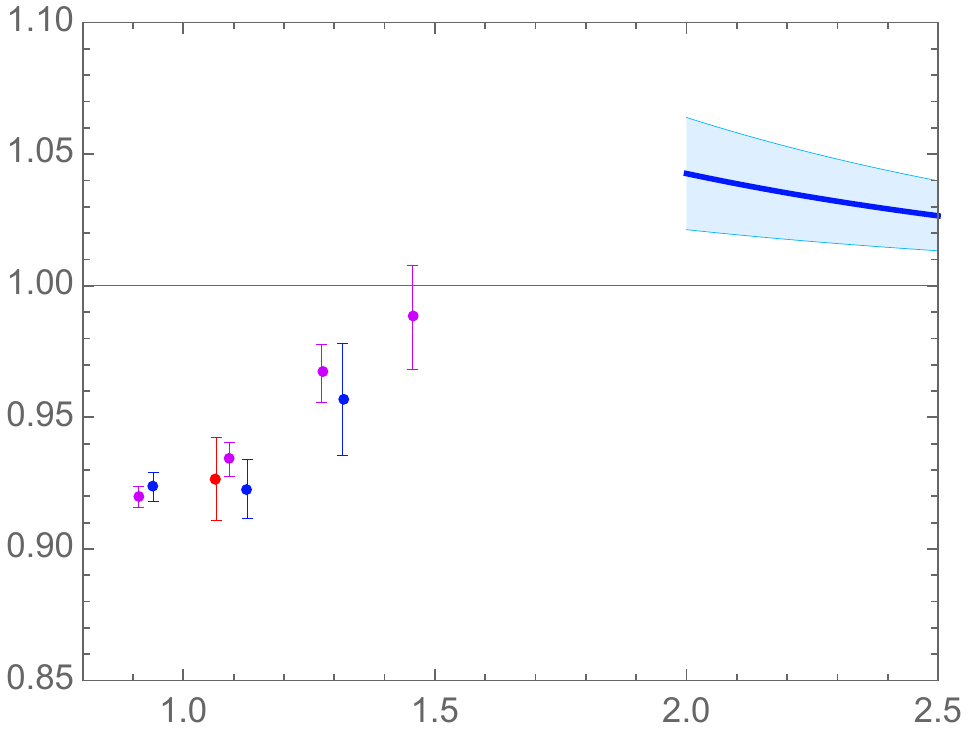}\\
$t$ [fm]
\caption{Lattice plateau estimates for the axial charge normalized by the experimental value. Lattice data from ETMC (blue), NME (magenta) and RQCD (red). The ChPT result of fig.\ \ref{fig:resultsall} is also shown for $t\ge 2$ fm by the blue solid line together with a 50\% error band as a naive estimate for higher order corrections.}
\label{fig:compgA}
\end{figure}

Figure \ref{fig:compgA} shows the renormalized plateau estimates obtained by  ETMC \cite{Abdel-Rehim:2015owa,Alexandrou:PC}, NME \cite{Yoon:2016jzj,Gupta:Lattice17} and RQCD \cite{Bali:2014nma}, divided by the experimental value $g_{A,{\rm exp}}$ \cite{Olive:2016xmw}.\footnote{The ETMC results shown in fig.\ \ref{fig:compgA} are based on larger statistics compared to the ones given in Ref.\ \cite{Abdel-Rehim:2015owa}. They have smaller error bars but are still preliminary \cite{Alexandrou:PC}. I thank C.\ Alexandrou and C.\ Kallidonis for sharing these results. I also thank R.\ Gupta, Y.-C.\ Jang and B.\ Yoon for sending me numerical results published in Ref.\ \cite{Yoon:2016jzj}. } The ETMC results (blue symbols) were obtained with $N_f=2$ twisted mass fermions with $M_{\pi}\approx 130$ MeV, $M_{\pi}L\approx 3$ and $a\approx 0.093$. The NME results (magenta symbols) were generated with $N_f=2+1$ Wilson-clover fermions  with $M_{\pi}\approx 165$ MeV, $M_{\pi}L\approx 3.7$ and $a\approx 0.09$. The remaining data point (red symbol) from RQCD was obtained with $N_f=2$ improved Wilson fermions with $M_{\pi} \approx 150$ MeV, $M_{\pi}L\approx 3.5$ and $ a\approx 0.071$ fm. More details about the simulation setup are given in the original papers. 

The plateau estimates shown in fig.\ \ref{fig:compgA} were obtained for source-sink separations between 0.9 and 1.5 fm. They are all below the experimental value although the NME result at $t\approx1.5$ fm agrees with $g_{A,{\rm exp}}$ within the statistical error. 

The ChPT prediction for the overestimation due to the $N\pi$ states is also shown for $t\ge 2$ fm. A naive guess for the higher order corrections is also included in form of a 50\% error band. Apparently, the lattice data is not in conflict with the ChPT result. It seems very plausible that the lattice data will  connect smoothly to the ChPT prediction when $t$ is increased. Whether this indeed happens needs to be checked. This requires lattice calculations of $g_A$ at $t$ larger than 1.5 fm with a statistical error of a few percent.\footnote{In addition one should also check that the underestimation of the axial charge at small source-sink separations persists in the continuum limit.} 

If indeed realized, such a scenario can be quite misleading in practice. Since the plateau estimate approaches the experimental value at some $t$ well before the asymptotic region is reached, one might be tempted to stop simulating at larger source-sink separations. In that case one reproduces the correct experimental value $g_{A,{\rm exp}}$, but for the wrong reason. The excited-state contributions are not small because $t$ is sufficiently large to be in the asymptotic regime. Instead, various excited-state contributions are still sizeable but accidentally cancel each other. 

A concrete model for such a scenario was suggested by M.\ Hansen and H.\ Meyer in Ref.\ \cite{Hansen:2016qoz}. Based on plausible assumptions concerning the higher order corrections to the LO ChPT results, the high-momentum $N\pi$ states with energies larger than about 1.5 $M_N$ contribute {\em negatively} to the plateau estimate. Summing up the total $N\pi$ contamination the positive contribution from the low-momentum states is overcompensated by the contribution from the high-momentum states, leading to an underestimation of the axial charge for source-sink separations below about $1.5$ fm, in agreement with lattice results. 

Whether the high-momentum $N\pi$ states are indeed responsible for the compensation of the low-momentum ones needs to be corroborated. Other excited states (e.g.\ $N\pi\pi$ and $\Delta\pi$ states) are expected to play a non-negligible role too for source-sink separations below 2 fm. In any case, the model in Ref.\ \cite{Hansen:2016qoz} supports the expectation that source-sink separations $t\gtrsim2$ fm are needed for the LO ChPT results to be applicable. 

\begin{figure}[p]
\centering
$\langle x\rangle_{u-d,{\rm plat}}(t)/\langle x\rangle_{u-d,{\rm phen}}$ \\
\includegraphics[scale=0.85]{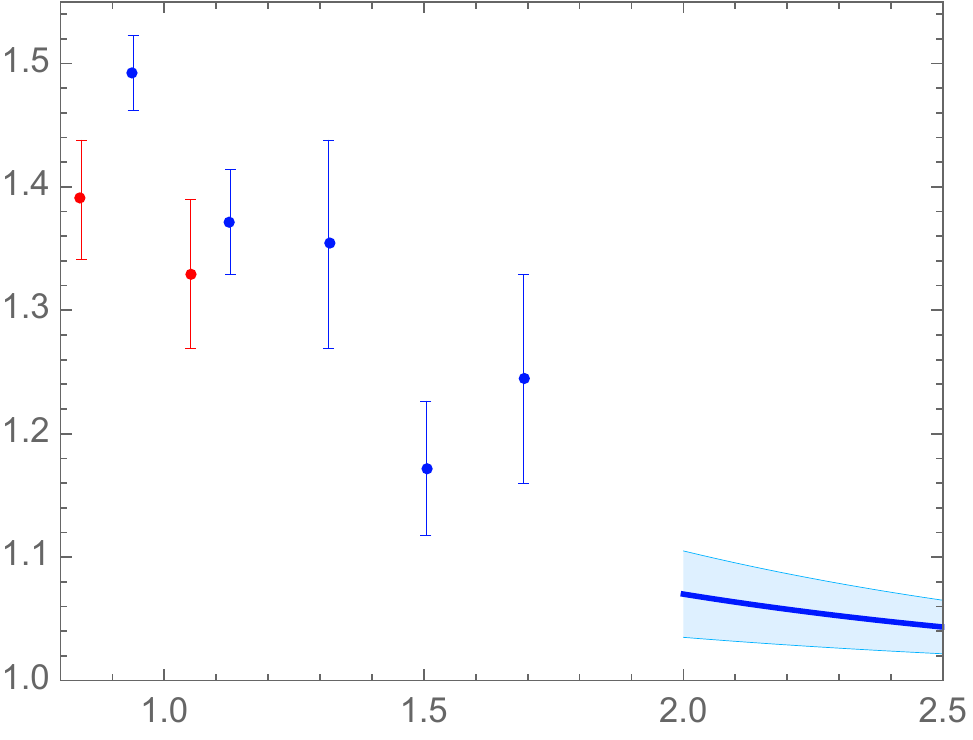}\\
$t$ [fm] \\[5ex]
$\langle x\rangle_{\Delta u-\Delta d,{\rm plat}}(t)/\langle x\rangle_{\Delta u-\Delta d,{\rm phen}}$  \\
\includegraphics[scale=0.85]{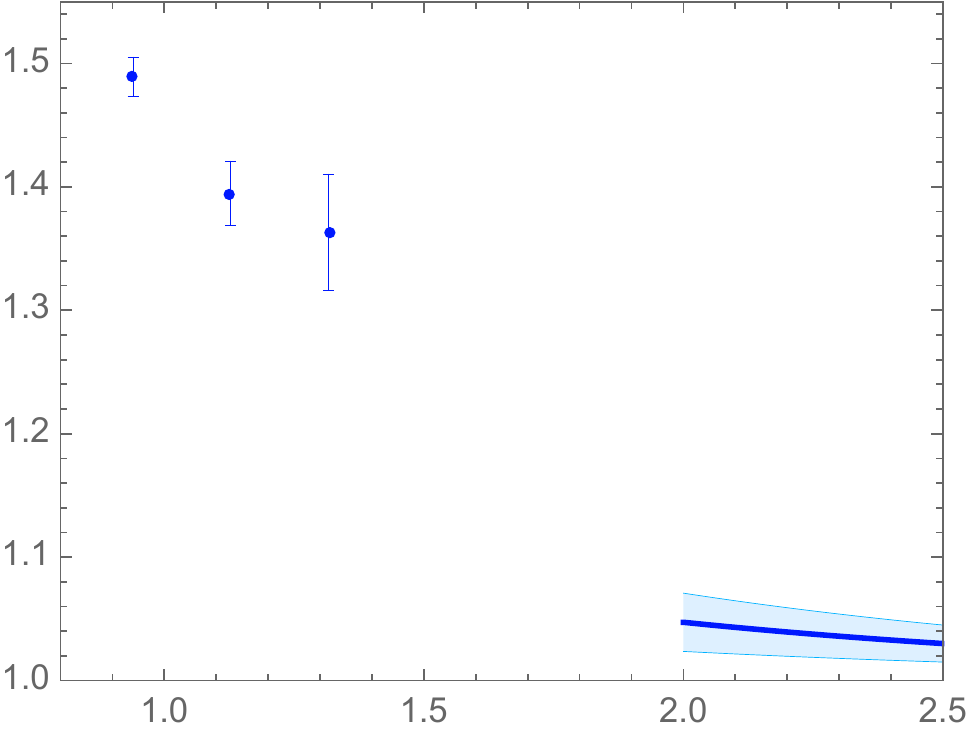}\\
$t$ [fm]
\caption{Lattice plateau estimates for the average momentum fraction (upper panel) and the helicity moment (lower panel) normalized by their phenomenological values.  Lattice data from ETMC (blue) and RQCD (red). The ChPT result of fig.\ \ref{fig:resultsall} is also shown for $t\ge 2$ fm together with a 50\% error band as a naive estimate for higher order corrections.}
\label{fig:compavmom}
\end{figure}

Fig.\ \ref{fig:compavmom}  shows the plateau estimates of ETMC \cite{Abdel-Rehim:2015owa,Alexandrou:PC} and RQCD \cite{Bali:2014gha} for the average quark momentum fraction $\mf$ and the helicity moment $\hm$, normalized by their phenomenological values given in \cite{Alekhin:2012ig,Blumlein:2010rn}. For both observables the plateau estimates overestimate the phenomenological value. The absolute discrepancy is significantly larger than for the axial charge (note the different scales in fig.\ \ref{fig:compavmom}).  The statistical errors are rather large, but the plots suggest that the plateau estimates decrease as the source-sink separation increases. Compared with the results for the axial charge it is here much simpler to imagine a simple monotonic decrease until contact with ChPT can be made at source sink-separations between 2 and 2.5 fm. 

In case of the scalar and tensor charges  as well as the transversity moment we do not have experimental values at our disposal. Thus, comparisons between the lattice plateau estimates and the ChPT predictions cannot be done. However, there is no reason to believe that the asymptotic region is reached at significantly smaller source-sink separations for these observables. In fact, the absence of experimental results  
elevates the role ChPT can play as a guide for the systematic uncertainties due the $N\pi$ states, provided one is in the asymptotic regime where ChPT can be applied.
 
The main conclusion we can draw from our comparison is that present lattice data at source-sink separations $t\lesssim 1.5$ fm is still far away from the asymptotic regime where the lowest lying $N\pi$ states are responsible for the dominant excited-state contribution in the plateau estimates. Lattice data at significantly larger source-sink separations (and with sufficiently small statistical errors) are needed to make contact with the asymptotic region where ChPT can be applied. 

\section{Miscellaneous comments}

An alternative way to compute the charges and pdf moments is provided by the summation method \cite{Maiani:1987by,Capitani:2012gj}. 
It starts from the ratio $R_X(t,t')$ and computes the sum over all operator insertion times between source and sink. In case of the axial charge the result of the sum is of the form
\begin{equation}\label{sumestimate}
\sum_{t'=0}^t R_A(t,t') = t[g_A + {\rm O}(e^{-\Delta E_n t})] + \dots\,.
\end{equation}
The slope $[\ldots]$ is called the summation estimate of the axial charge. For finite $t$ it differs by exponentially suppressed excited state contributions from $g_A$. 

Using the ChPT results for the ratios in eq.\ \pref{sumestimate} the $N\pi$-state contribution to the summation estimate is easily computed. However, the sum on the left hand side involves the 3pt function at short distances when $t'$ approaches either source or sink. In these cases the ChPT results do not properly capture the full 3pt function and the ratio.

A possible remedy for this limitation would be to keep all time separations large by summing over a central subinterval only. This, however, requires even larger source-sink separations than those needed for the plateau method. Based on our results we would expect to need a minimal separation of about 1 fm for $t-t'$ and $t'$. In addition we need a non-zero time interval to sum over. Both requirements imply minimal source-sink separations of about 2.5 fm if not larger. Note that this is not a requirement for the summation method per se, but for ChPT to have something to say about it.

\begin{figure}[t]
\centering
$M_{\rm eff}(t)/M_N$\\
\includegraphics[scale=0.85]{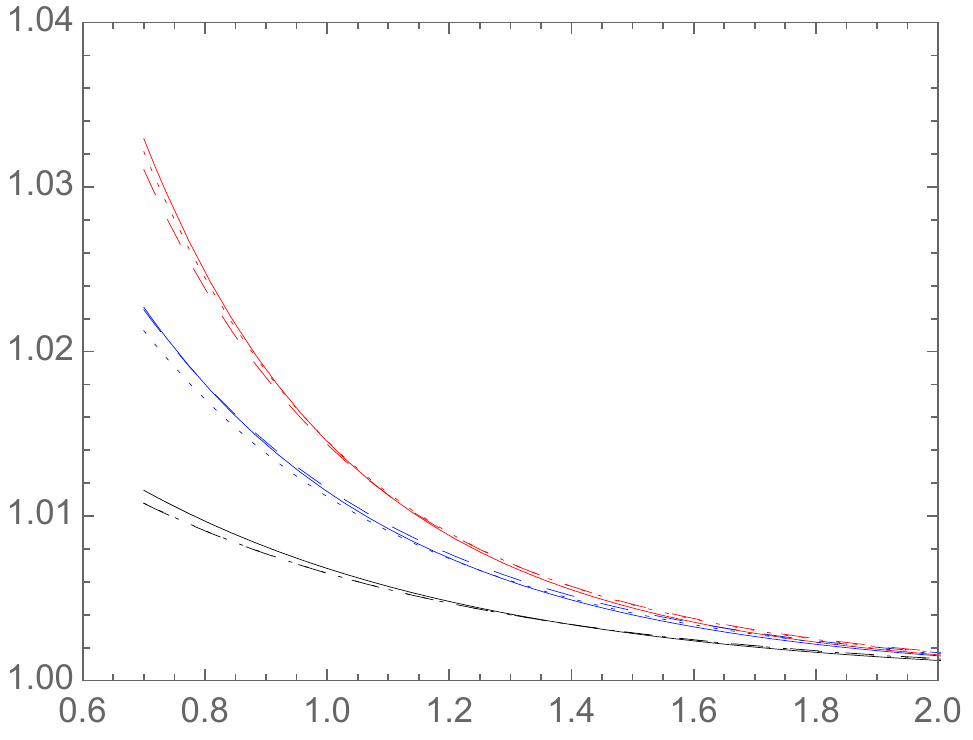}\\
$t$ [fm]
\caption{The ratio $M_{N,{\rm eff}}(t)/M_N$ as a function of the source sink separation $t$. Results are shown for $M_{\pi}L=4$ (solid lines),  $M_{\pi}L=5$ (dotted lines) and $M_{\pi}L=6$ (dashed lines). Different colors distinguish between three different upper energy bounds for the $N\pi$ states taken into account (see main text).}
\label{fig:effmass}
\end{figure}

For the computation of the various ratios we had to compute the 2pt function. It is interesting to also look at the $N\pi$ contribution to the effective nucleon mass, c.f.\ eg.\ \pref{Meff}. Fig.\ \ref{fig:effmass} shows the ratio $M_{N,{\rm eff}}(t)/M_N$ as a function of the source-sink separation $t$. 
Without the excited state contribution this ratio would be equal to 1, any deviation from this value stems from the $N\pi$ contribution. As in figure \ref{fig:axialcharge}, results are shown for the three different lattice sizes (solid, dashed and dotted lines)  and the three different energies $E_{N\pi,{n_{\rm max}}}$ (black, blue and red lines) considered before.

Qualitatively figs.\ \ref{fig:axialcharge} and \ref{fig:effmass} look very similar. Only a very small FV dependence is visible, provided the number of states are adjusted according to the energy interval for the $N\pi$ states taken into account for the $N\pi$ contribution. The larger $t$ the more dominates the contribution of the low-momentum $N\pi$ states (black curves). The actual time $t$ where this happens, however, is much smaller compared to the result for the axial charge. Even more importantly, the $N\pi$ state contamination in the effective mass is significantly smaller. Even allowing for an error as big as 100\% for the $N\pi$ contribution the overestimation of the nucleon mass is smaller than 2\% at source-sink separations of about  1.2 fm and larger. This contamination can be ignored unless the lattice data have statistical errors at the sub-percent level.

Note that the significantly larger excited-state contamination in the charges and moments stems from the contributions associated with the time separations $t-t'$ and $t'$. For this reason the $N\pi$ contribution at a given time $t$ in fig.\ \ref{fig:effmass} should be compared the one at $2t$ in fig.\ \ref{fig:axialcharge}. In that case the $N\pi$ contribution in both observables is roughly of the same size.\footnote{The factors $\Delta E_n/M_N$ in $M_{\rm eff}/M_N$, c.f.\ eq.\ \pref{Meff}, causes an additional suppression of the excited-state contribution in the effective mass.} 

The conclusions we have drawn are based on the LO results for the $N\pi$-state contribution. This contribution dominates for large time separations in physical point simulations. A calculation of the NLO corrections is certainly desirable, mainly to obtain firmer error estimates for the LO results. 
In addition, it needs to be checked that contributions of other excited states are small and do not change qualitatively the results we have found. Other multi-hadron-state contributions have their origin in three-particle $N\pi\pi$ states and two-particle $\Delta \pi$ states. ChPT calculations of these contributions are essentially analogous to the calculation of the $N\pi$ contribution and in principle straightforward to perform. 
 
\section{Conclusions}

Physical point simulations eliminate the systematic uncertainties associated with a chiral extrapolation. Excited-state effects due to multi-hadron-states involving one or more pions, on the other hand, become more pronounced for pion masses as light as in Nature. Using Baryon ChPT we have computed the two-particle $N\pi$ state contamination in the plateau estimates of various nucleon charges and pdf moments. This particular excited-state contamination leads to an overestimation at the 5-10\% level for source-sink separations of about 2 fm. This is uncomfortably large and cannot be not ignored if results with percent precision are the goal of lattice calculations. 

Even more troublesome is that lattice calculations of the charges and moments with source-sink separations of 2 fm and larger are out of reach with present simulation techniques. So far most lattice results have been obtained with source-sink separations of 1.5 fm and smaller. At time separations that small one expects excited states other than $N\pi$ states to contribute a sizeable excited-state contamination too. 

In case of the axial charge the lattice data together with the ChPT result strongly suggests a non-monotonic $t$ dependence of the plateau estimate. The lattice data underestimate the experimental value for $t\lesssim 1.5$ fm while ChPT predicts an overestimation for larger time separations. This peculiar behavior, neither seen for the average quark momentum fraction nor the helicity moment, can be attributed to other excited states that contribute a negative contamination to the plateau estimate. In this case we expect a cancellation of the various excited-state contributions for some source-sink separation and an accidental agreement between the lattice result and the experimental value for the axial charge. Given the benchmark character of $g_A$ it thus seems vital to carefully check the excited-state contamination in this quantity with lattice simulations at source-sink separations larger than 1.5 fm.  

The main conclusion we can draw from our ChPT calculation is that physical point simulations require significantly larger source-sink separations than those with heavier pion masses. Simulations with  larger time separations require new simulation techniques to overcome the notorious signal-to-noise problem. Some ideas in that respect have been recently proposed \cite{Ce:2016idq,Ce:2016ajy,Wagman:2017xfh} and presented at this conference \cite{Giusti:Lat17,Wagman:Lat17}, but remain to be tested in actual lattice calculations of the nucleon correlation functions considered here.

Alternatively one may resort to the variational method \cite{Luscher:1990ck,Blossier:2009kd} to get a handle on the nucleon matrix elements at smaller time separations. The variational method can be used provided interpolating fields for the $N\pi$ states are taken into account \cite{Lang:2012db,Kiratidis:2015vpa,Lang:2016hnn,Kiratidis:2016hda}. In this approach too the numerical cost grows significantly since the number of Wick contractions involved in computing the correlation functions grows rapidly once interpolating fields for multi-particle states are used \cite{Lang:2016hnn}. 

\section*{\small Acknowledgments}
Communication with C.~Alexandrou, R.~Gupta, Y.-C.~Jang, M.~Kallidonis and R.~Sommer is gratefully acknowledged. 
This work was supported by the Japan Society for the Promotion of Science (JSPS) with an Invitation Fellowship for Research in Japan (ID No. L16520), and by the German Research Foundation (DFG), Grant ID BA 3494/2-1.

%\bibliography{lattice2017}

%%%%%%%%%%%%%%%%%%%%%%%%%%%%%%%%%%%%%%%%%%%%%%%%%%%%%%%%
\end{document}